\documentclass[10pt]{iopart}
\usepackage{iopams}  
\usepackage{cite}
\usepackage{hyperref}
\expandafter\let\csname equation*\endcsname\relax
\expandafter\let\csname endequation*\endcsname\relax
\usepackage{amsmath}
\usepackage{txfonts}
\usepackage{xcolor}
\usepackage{bm}
\begin{document}

\title{Stronger EPR-steering criterion based on inferred Schr\"odinger-Robertson uncertainty relation}

\author{Laxmi Prasad Naik$^{1,}$$^2$, Rakesh Mohan Das$^3$ and Prasanta K. Panigrahi$^2$}

\address{$^1$ Indian Institute of Science Education And Research Kolkata, Mohanpur, Nadia - 741 246, West Bengal, India}
\address{$^2$ Indian Institute of Technology Delhi, Hauz Khas - 1100016, New Delhi, India}
\address{$^3$ Kalinga Institute of Industrial Technology, Bhubaneswar - 751024, Odisha, India}
\ead{laxmiprasadnaik5897@gmail.com,rakesh.dasfpy@kiit.ac.in and pprasanta@iiserkol.ac.in}
\vspace{10pt}

\begin{abstract}
Steering is one of the three in-equivalent forms of nonlocal correlations intermediate between Bell nonlocality and entanglement. Schr\"odinger-Robertson uncertainty relation (SRUR), has been widely used to detect entanglement and steering. However, the steering criterion in earlier works, based on SRUR, did not involve complete inferred-variance uncertainty relation. In this paper, by considering the local hidden state model and Reid’s formalism, we derive a complete inferred-variance EPR-steering criterion based on SRUR in the bipartite scenario. Furthermore, we check the effectiveness of our steering criterion with discrete variable bipartite two-qubit and two-qutrit isotropic states.
\end{abstract}

%
\vspace{2pc}
\ioptwocol
\section{Introduction}
EPR-steering is a nonlocal correlation intermediate between Bell nonlocality and quantum entanglement \cite{einstein_1935,einstein_1936,wiseman2007steering}. It is the ability to remotely affect or \emph{steer} subsystems of a shared entangled quantum state by an arbitrary choice of local measurements on the subsystems without violating the no-signaling principle \cite{simon2001no}. Wiseman \textit{et al.} gave an operational definition of steering as a task between Alice and Bob. Alice prepares an entangled state and sends one part to Bob. Here, Bob does not trust Alice, and by performing local measurements, she has to convince him that the state is entangled \cite{wiseman2007steering}. If Bob's steered quantum state cannot be explained by a local hidden state (LHS) model, then the state is said to exhibit steering. In contrast to Bell's nonlocality and entanglement, steering demonstrates asymmetric behaviour in which one party can steer the other party, but vice versa is not always permitted \cite{midgley2010asymmetric,bowles2014one,bowles2016sufficient,reid2013monogamy}. Moreover, not every entangled state exhibits steering, and not every steerable state violates Bell inequality \cite{wiseman2007steering}.
EPR-steering has a wide range of applications in many quantum information processing tasks, e.g., in one-sided device-independent quantum key distribution \cite{branciard2012one,gehring2015implementation,walk2016experimental}, quantum networking tasks \cite{Huang2018SecuringQN,Armstrong2014MultipartiteES,Cavalcanti2014DetectionOE}, subchannel discrimination \cite{Piani2015NecessaryAS,Chen2016NaturalFF, Sun2018DemonstrationOE}, quantum secret sharing \cite{Kogias2016UnconditionalSO, Xiang2016MultipartiteGS}, quantum teleportation \textcolor{blue}{\cite{fan2022quantum}}, randomness certification \cite{passaro2015optimal,curchod2017unbounded,Mattar2016ExperimentalME}, and random number generation \cite{joch2022certified} to mention a few. Recently, it has also been demonstrated to be a useful resource in noisy and lossy quantum network systems \cite{Qu2021RetrievingHQ,Srivastav2022QuickQS}.

Effective detection of steering exhibited by quantum states is crucial to realise applications of steerable quantum states. Uncertainty relations (UR) can be experimentally verified because it involves measurement of observables. There have been many works involving the use of UR's for detection of entanglement \cite{ViolationofentanglementPhysRevA.68.032103} for discrete variables and continuous variables \cite{Einstein-Podolsky-RosenuncertaintylimitsforbipartitemultimodestatesPhysRevA.103.062224}. Assuming that the description of quantum mechanics is correct, EPR's condition of locality and sufficient condition of reality are satisfied, UR's become an important tool for determining steering criteria. Many criteria in this direction, e.g., using the Heisenberg uncertainty relation (HUR) \cite{reid1989demonstration} and later involving a broader class of uncertainty relations have been proposed \cite{bialynicki1975uncertainty, Deutsch1983UncertaintyIQ,Chowdhury2013EinsteinPodolskyRosenSU,schneeloch2013einstein}. Additionally under different measurement scenarios more optimal steering criteria \cite{pramanik2014fine, maity2017tighter} were obtained using fine-grained \cite{Oppenheim2010TheUP, Chowdhury2015StrongerSC} and sum uncertainty relations \cite{Maccone2014StrongerUR, maity2017tighter}.

The criterion for experimental demonstration of steering was first proposed by Reid \cite{reid1989demonstration}, which is based on inferred-variances. Recently, a steering criterion using Schr\"odinger-Robertson uncertainty relation (SRUR) was also proposed. However, these earlier works used only Reid criterion and did not adopt the LHS model, hence did not use inferred-means in the lower bound \cite{sasmal2018tighter}. A recent work involves inferred-variance based product and sum uncertainty relations in the presence of entanglement \cite{bagchi2022inferred}. We aim to derive a steering criterion based on SRUR involving inferred-means and inferred-variances following up the analysis in \cite{cavalcanti2009experimental}.

SRUR is a generalized relation and reduces to HUR when the covariance between two operators is zero, which is not always true. Hence SRUR provides a stronger bound than HUR. Classically the term $\langle x^{p}y^{p} \rangle - \langle x^{p} \rangle \langle y^{p} \rangle$ indicates the degree of statistical correlation or dependency between any two random variables $x$ and $y$. The covariance term in SRUR is $\frac{ \{ \langle \hat{B}_{1},\hat{B}_{2} \rangle\} }{2} -\langle\hat{B}_{1}\rangle\langle\hat{B}_{2}\rangle$, where $\hat{B}_{1}$ and $\hat{B}_{2}$ are any two hermitian operators is termed as “quantum correlation coefficient”. If the two operators are statistically independent, then the covariance term vanishes. SRUR describes an inevitable relationship among the lowest order moments of the operators $\hat{B}_{1}$ and $\hat{B}_{2}$ i.e., $\{\hat{B}_{1}^{q},\hat{B}_{2}^{q}\}: q = 1,2$, for which one needs to have access to the statistics or the expectation values of the two other operators viz., the anticommutator $\{\hat{B}_{1},\hat{B}_{2}\}$ and the commutator $[\hat{B}_{1},\hat{B}_{2}]$ which is obtained from two different and identical runs of the experiment.

Additionally, there is no end to the list of observables, $\{\hat{B}_{1},\hat{B}_{2},\hat{B}_{3},\hat{B}_{4},\hat{B}_{5},....\}$,  of which one could construct moments of all orders and construct a generalized version of SRUR. A generalized version of SRUR involving higher order moments can also be constructed. Quantum mechanics can, in its entirety,
be portrayed as a “theory of interactive moments.” This perspective emphasizes the fundamental role and significance of higher order moments in capturing the interactive nature of quantum phenomena of higher dimensional states.
It is shown that a stricter entanglement condition is provided by the SRUR, using positive partial transpose for non-Gaussian entangled 
states \cite{NhaPhysRevA.76.014305,goswami2017uncertainty} and  using covariance matrix entanglement criterion for Gaussian entangled states \cite{SimonPhysRevLett.84.2726}. This has been later generalized by V Tripathi \emph{et al.} \cite{Tripathi_2020}.

In the next section, we briefly discuss steering and the EPR-Reid criterion. We illustrate how EPR-Reid criterion helps in detection of steering. In Sec.3, based on SRUR, we derive a steering criterion to detect steerability of states by incorporating the EPR-Reid criterion and the LHS model. We then check the efficiency of the steering criterion in Sec.4, using it on two-qubit and two-qutrit isotropic states for which corresponding steering inequalities were obtained. We conclude with a discussion on the strength and limitations of the steering criterion, along with the future scope of our work. All the detailed derivation is included in the appendix.
\section{Preliminaries}
\subsection{EPR-Steering}\label{EPR Steering}
Consider a general unfactorizable bipartite pure state shared by two distant parties, Alice and Bob,
\begin{eqnarray}\label{eqn1}
    |\Psi\rangle = \sum_{n} c_n|u_n\rangle|v_n\rangle = \sum_{n} d_n|\psi_n\rangle|\phi_n\rangle
\end{eqnarray}
where, $\{|u_n \rangle \}(\{|\psi_n\rangle\})$ and $\{|v_n \rangle \}(\{|\phi_n\rangle\})$ denote two different orthonormal bases in Alice’s and
Bob’s system, respectively. This property of inseparability is called entanglement, which is one of the most useful resources in quantum information processing that has been studied extensively in the literature \cite{horodecki,Bhaskara_QINP_2017,roy2021geometric,mahanti2022classification,mishra2022geometric}. In steering scenario, Alice chooses to measure in the $|u_n\rangle\left(|\psi_n\rangle\right)$ basis, then Bob's state is projected into $|v_n\rangle (|\phi_n\rangle)$ basis. The ability of Alice to influence (steer) Bob’s state, nonlocally was termed as \textit{steering} by Schr\"odinger \cite{einstein1935can,einstein_1935,einstein_1936}.

Alice and Bob share an entangled quantum state, described by density matrix $\hat{\rho}$. The generalised local measurements of Alice and Bob are denoted by ${\hat{M}_{a|A}}$ and ${\hat{M}_{b|B}}$ ($M_{a(b)|A(B)} \geq 0, \sum_{a(b)}M_{a(b)|A(B)} = 1 \hspace{3mm}\forall A(B)$, where $a$ and $b$ denote the outcomes corresponding to the measurement operators $\hat{M}_{a|A}$ and $\hat{M}_{b|B}$. $A$ and $B$ are Alice's and Bob's measurement settings, respectively. The quantum probability of their joint measurement is given as
\begin{eqnarray}\label{eqn2}
    P(a,b) = \textnormal{Tr}\left[\hat{\rho}(\hat{M}_{a|A} \otimes \hat{M}_{b|B})\right]
\end{eqnarray}
where, $P(a,b)$ is the joint probability of obtaining outcomes $a$ and $b$. Alice fails to steer Bob's state, if and only if for all the measurements $\hat{M}_{a(b)|A(B)}$, the joint probability distribution $P(a,b)$ for Alice’s and Bob’s measurements can be explained using an LHS model for Bob and a local hidden variable (LHV) model for Alice. The joint probability distributions can be written as
\begin{eqnarray}\label{eqn3}
    P(a,b) = \sum_{\eta}p(\eta)P(a|\eta)P_{Q}(b|\eta)
\end{eqnarray}
where $\eta$ is a local hidden variable having probability distribution $p(\eta)$, satisfying $p(\eta) \geq 0$ and $\sum_{\eta}p(\eta) = 1$. $P(a|\eta)$ is the probability distribution for outcome $a$ determined by the local hidden variable $\eta$ and $P_{Q}$ is the quantum probability distribution for outcome $b$; $P_{Q}(b|\eta)= \textnormal{Tr}_{B}[\hat{M}_{b|B}\hat{\rho}_{\eta}]$ ($Q$ stands for quantum), correspond to a local hidden quantum state described by $\hat{\rho}_{\eta}$, which is unaffected by local measurements of Alice. The use of LHS to explain steering is a clear implication of the consistency of EPR's condition of locality. Any constraint that can be obtained obeying Eq.~\eref{eqn3} will form an \emph{EPR-steering criterion} violation of which will demonstrate steering. The joint probability distribution and the state is said to admit an LHS model if Eq.~\eref{eqn2} can be expressed having a decomposition of the form of Eq.~\eref{eqn3} for all the choice of Alice's and Bob's measurements respectively.

It is important to note that Alice and Bob perform single measurements in each run of the experiment on the identically prepared states. And, Alice arbitrarily chooses to perform different measurement settings in each of the runs of the experiment, i.e., $P(A_{i}|A_{j}) = P(A_{i})$ and $P(A_{j}|A_{i}) = P(A_{j})$ where $A_{i}$ and $A_{j}$ are Alice's measurement settings. The probability of the result of a measurement is independent of the other measurement results. Mathematically, the probability $P(a_{i}|a_{j}) = P(a_{i})$ and $P(a_{j}|a_{i}) = P(a_{j})$ where $a_{i}$ and $a_{j}$ are the results which correspond to different single measurements performed in each run by Alice.

The interpretation can be given in terms of Bob's conditional state i.e. Bob's state assemblages. Prior to all experiments, Bob asks Alice to announce the set of possible ensembles of states into which Alice would like to project Bob's state into i.e., $E^{A}: \forall A $. For every run of the experiment, Alice prepares an entangled state and sends one part to Bob. Bob randomly picks an ensemble $E^{A}$ from the announced set and asks Alice to prepare it. Alice performs the measurement $A$ and announces to Bob about his collapsed state $\hat{\sigma}^{A}_{a}$ corresponding to the result $a$ that she has prepared. This experiment is performed over many runs and Bob confirms the probability $\textnormal{Tr} [\hat{\sigma}^{A}_{a}]$ of its occurrence.

Alice can cheat Bob by adopting a strategy. In this strategy, instead of sending one part of an entangled state to Bob, Alice picks a state at random from some prior ensemble of states $\textnormal{R} = \{p_{\eta}\hat{\sigma}_{\eta}\}$ with $\hat{\sigma} = \sum_{\eta}p_{\eta}\hat{\sigma}_{\eta}$. Now, Alice announces to Bob about the state that is prepared from her knowledge of $\eta$ and any stochastic map from $\eta$ to $a$, where $a$ is one of the results of the measurement $A$, which Bob had picked from the announced ensemble. Alice will fail to convince Bob that the state sent is entangled if and only if, for all of Alice's choice of measurements, $\hat{A}$ and for all eigenvalues of $\hat{A}$, one can find an ensemble R of states with probability distribution $p_{\eta}$ and a stochastic map $p(a|A,\eta)$ from $\eta$ to $a$ such that the Bob's state can be explained by a local hidden state model,
\begin{eqnarray}\label{eqn4}
    \hat{\sigma}_{a}^{A} = \sum_{\eta}p_{\eta}p(a|A,\eta)\hat{\sigma}_{\eta}
\end{eqnarray}
For the state assemblages of Bob, if Alice cannot find an ensemble R, i.e., a local hidden state model $\hat{\sigma}_{\eta}$ and a stochastic map $p(a|A,\eta)$, then the state is steerable. Alice cannot affect Bob's unconditioned state $\textnormal{Tr}_{A}[\hat{\rho}]$, because that would violate superluminal communication \cite{simon2001no}.
In this strategy, we observe that in each run of the experiment, while Bob picks one of the ensembles $E^{A}$ announced by Alice and asks her to announce back to him about his collapsed state. Alice performs this step of the experiment through a stochastic map from $\eta$ to $a$, corresponding to one of the results, $a$ of the measurement $A$. The choice of which outcome of the measurement should be mapped from $\eta$ to $a$, is completely arbitrary and random, i.e.. $P(a_{i}|a_{j}) = P(a_{i})$, $\forall a_{i},a_{j}$ corresponding to different results of Alice's different measurements.

In a local hidden state model, the choices of measurements of Alice to infer the values of the measurements of Bob are arbitrary. The reason is, Alice’s probabilities are allowed to depend arbitrarily on the variables $\eta$. The local hidden state model description of steering uses the property of states. However, in an experimental situation, one is not concerned about what type of state is used to demonstrate steering rather we only rely on the measured data. As in the case of bell nonlocality and entanglement detection techniques we rely on the measured data only rather than the state properties i.e., local hidden variables and local hidden states, respectively. Therefore an experimental EPR-steering criterion should not depend on any assumption about the type of state being prepared rather depend only on the measured data.

Alice can attempt to infer different outcomes of Bob corresponding to different observables and assuming the LHV-LHS model, since Bob's state corresponds to a local hidden quantum state, uncertainty relations can be used for Bob's measurements. This was first realized by Reid \cite{reid1989demonstration}, who proposed an experimental EPR-steering criterion using HUR for continuous variable systems. We aim to derive an EPR-steering criterion using SRUR by incorporating LHS model and EPR-Reid criterion.

\subsection{EPR-Reid criterion}\label{EPR Reid}
Reid proposed a modified version of EPR's sufficient condition of reality which states that if without disturbing a system in any way one can predict the value of a physical quantity with some specified uncertainty then there exists a stochastic element of physical reality which determines this physical quantity with atmost that specific uncertainty, called as \emph{Reid's extension of EPR's sufficient condition of reality}. This is attributed to the intrinsic stochastic nature exhibited in the preparation and detection of quantum states \cite{reid1989demonstration,cavalcanti2009experimental}.

Consider two parties Alice and Bob sharing an entangled state. Now Alice makes a local measurement $\hat{Y}$ and makes an estimate $\hat{X}^{est}(\hat{Y})$ for the result of Bob's measurement $\hat{X}$ by inferring from the outcomes of her own measurement $\hat{Y}$. The idea of estimation is implemented to incorporate EPR's sufficient condition of reality. Therefore the average inferred-variance of $\hat{X}$ for an estimate $\hat{X}^{est}(\hat{Y})$ is given by
\begin{eqnarray}\label{eqn5}
    \Delta_{inf}^{2}\hat{X}^{2} = \left\langle\left(\hat{X} - \left\langle\hat{X}^{est}(\hat{Y})\right\rangle\right)^{2}\right\rangle.
\end{eqnarray}
Alice's estimate for Bob's measurement is given by $\hat{X}^{est}(\hat{Y}) = g\hat{Y}$, where the choice of $g$ should be such that it gives the minimum error, i.e., $g = \frac{\langle\hat{X}\otimes\hat{Y}\rangle}{{\langle \hat{Y}^{2} \rangle}}$ gives the optimal inferred-variance. Using EPR's condition of locality Reid's extension of EPR's sufficient condition for reality and completeness of quantum mechanics (a limit on the product of inferred-variances) based on HUR for two noncommuting quadrature phase observables $\hat{X}_{1}$ and $\hat{X}_{2}$ on Bob's side is \cite{reid1989demonstration} given as 
\begin{eqnarray}\label{eqn6}
\Delta_{inf}^{2}\hat{X}_{1}\Delta_{inf}^{2}\hat{X}_{2} \geq 1.
\end{eqnarray}
This is known as \emph{EPR-Reid criterion}. A state will show steering if Eq.~\eref{eqn6} is violated, which has also been verified experimentally \cite{cavalcanti2009experimental}.

\section{EPR-steering criterion using Schr\"odinger-Robertson uncertainty relation}
Our derivation of EPR-steering is based on the works of \cite{reid1989demonstration,cavalcanti2007uncertainty,cavalcanti2009experimental}. Here, we use a different notation for the outcomes of measurement. Consider the outcomes $A$ and $B$ corresponding to observables $\hat{A}$ and $\hat{B}$ for Alice's and Bob's measurements respectively. Alice and Bob perform measurements $\hat{A}$ and $\hat{B}$ respectively, and Alice tries to infer the outcome of Bob's measurement based on the result of her measurement. $B^{est}(A)$ is Alice's estimate of Bob's measurement outcome based on her outcome $A$.
Using the EPR-Reid criterion the inferred-variance is written as
\begin{eqnarray}\label{eqn7}
    \Delta_{inf}^{2}B = \left\langle\left(B - B^{est}(A)\right)^{2}\right\rangle.
\end{eqnarray}
The average is calculated by taking the average over all outcomes $A$ and $B$ by repeating the experiment over many runs, where each run involves a single pair of measurement on Alice's and Bob's side. The estimates involving different measurements is calculated following the EPR-Reid criterion. This scenario is similar to the detection of Bell nonlocality and entanglement. The inferred-variance $\Delta_{inf}^{2}B$ is minimized (optimized) when $B^{est}(A) = \langle B \rangle_{A}$. So the minimized inferred-variance $ \Delta^{2}_{min}B$ is as follows
\begin{eqnarray}\label{eqn8}
     \Delta^{2}_{min}B &= \left\langle\left(B - B^{est}(A)\right)^{2}\right\rangle = \sum_{A,B} P(A,B)\left(B - B^{est}(A)\right)^{2} \nonumber \\
                            &= \sum_{A} P(A) \sum_{B} P(B|A)\left(B - \langle B \rangle_{A}\right)^{2} \nonumber \\
                            &= \sum_{A} P(A) \Delta^{2} (B|A).
\end{eqnarray}
where $\Delta^{2} (B|A)$ stands for the conditional variance of Bob's measurement outcome $B$ provided the outcome $A$ of Alice's measurement is known. So we have the following condition:
\begin{eqnarray}\label{eqn9}
    \Delta^{2}_{inf}B \geq \Delta^{2}_{min}B.
\end{eqnarray}
Assuming that the statistics of the experimental outcomes of Alice's and Bob's measurements can be described by an LHS model in Eq.\eref{eqn3}, the conditional probability distribution $P(B|A)$ can be written as
\begin{eqnarray}\label{eqn10}
    P(B|A) &= \frac{P(A,B)}{P(A)} = \sum_{\eta} \frac{P(\eta) P(A|\eta)}{P(A)} P_{Q}(B|\eta) \nonumber \\
                       &= \sum_{\eta} P(\eta|A)P_{Q}(B|\eta).
\end{eqnarray}
Here, $\eta$ is a classical random variable such that $P(\eta) \geq 0$ and $\sum_{\eta}P(\eta) = 1$. Moreover, we can observe that the basic essence of adopting the LHS model is statistical independence of probabilities which is one of the most important prescriptions in the LHV theory by Bell \cite{bell1964einstein}. If $P(u)$ is a classical probability distribution which has a convex decomposition, i.e., $P(u) = \sum_{v} P(v)P(u|v)$, then the variance, $\Delta^{2} u$ corresponding to the probability distribution $P(u)$ is bounded by the average of the variances $\Delta^{2}(u|v)$ over the conditional distribution $P(u|v)$, i.e., $\Delta^{2} u \geq \sum_{v}P(v)\Delta^{2}(u|v)$. Therefore, from Eq.~\eref{eqn9} the variance of the conditional measurement outcomes $B|A$ is given by 
\begin{eqnarray}\label{eqn11}
    \Delta^{2}(B|A) \geq \sum_{\eta}P(\eta|A)\Delta^{2}_{Q}(B|\eta)
\end{eqnarray}
where the variance $\Delta^{2}_{Q}(B|\eta)$ is calculated using the conditional quantum probability distribution, $P_{Q}(B|\eta) = \textnormal{Tr}[\hat{B}\hat{\rho}_{\eta}]$. The average of the measurement operator $\hat{B}$ specified by its outcome $B$ is calculated corresponding to a local quantum hidden state described by $\hat{\rho}_{\eta}$. Therefore, using Eq.\eref{eqn11} the bound for $\Delta_{min}^{2}B$ is given as follows
\begin{eqnarray}\label{eqn12}
\Delta^{2}_{min}B &\geq \sum_{A} P(A) \Delta^{2} (B|A) \nonumber \\ &\geq \sum_{A} P(A) \sum_{\eta}P(\eta|A)\Delta^{2}_{Q}(B|\eta) \nonumber \\ &\geq \sum_{A,\eta} P(A,\eta)\Delta^{2}_{Q}(B|\eta) \nonumber \\ &\geq \sum_{\eta} P(\eta)\Delta^{2}_{Q}(B|\eta).
\end{eqnarray}
Consider Bob's arbitrary local measurement operators $\hat{B_{1}}$ and $\hat{B}_{2}$ with their corresponding outcomes $B_{1}$ and $ B_{2}$ respectively. These operators then satisfy the SRUR \cite{robertson1934indeterminacy}, i.e.,
\begin{eqnarray}\label{eqn13}
\langle\Delta^{2}\hat{B}_{1}\rangle_{\hat{\rho}} \langle\Delta^{2} B_2\rangle_{\hat{\rho}} \geq \frac{1}{4}\left|\langle[\hat{B}_{1},\hat{B}_{2}]\rangle_{\hat{\rho}}\right|^2 \nonumber + \\ \hspace{4cm} \left(\frac{\langle\{\hat{B}_1,\hat{B}_2\} \rangle}{2} - \langle \hat{B}_1\rangle \langle \hat{B}_2 \rangle \right)_{\hat{\rho}}^2.
\end{eqnarray}
where $\{\hat{B_1},\hat{B_2}\}$ is the anticommutator and $[\hat{B}_{1},\hat{B}_{2}]$ is the commutator of the two operators $\hat{B}_{1}$ and $\hat{B}_{2}$ respectively. $\langle\Delta_{Q}^{2}\hat{B_i}\rangle_{\hat{\rho}}$ is the variance and $\langle\hat{B_i}\rangle_{\hat{\rho}}$ is the average of the operator $\hat{B}_{i}$ calculated for a quantum state $\hat{\rho}$. The above equation can be written in terms of the outcomes of Bob, given by
\begin{eqnarray}\label{eqn14}
    \langle\Delta_{Q}^{2}B_{1}\rangle \langle\Delta_{Q}^{2} B_2\rangle \geq \frac{1}{4}\left|\langle[B_{1},B_{2}]\rangle_{Q}\right|^{2} + \nonumber \\ \hspace{4cm} \left(\frac{\langle\{\hat{B}_1,\hat{B}_2\} \rangle}{2} - \langle \hat{B}_1\rangle \langle \hat{B}_2 \rangle\right)_{Q}^2.
\end{eqnarray}
For any two vectors $\textbf{u}$ and $\textbf{v}$ in a linear vector space, the Cauchy-Schwartz inequality is given by
\begin{eqnarray}\label{eqn15}
 ||\textbf{u}||^{2}_{2}||\textbf{v}||^{2}_{2} \geq |\langle\textbf{u},\textbf{v}\rangle|^2   
\end{eqnarray}
where $||.||_{2}$ is the L2 norm, $\langle.\rangle$ is the inner product and $|.|$ is the modulus in the linear vector space.
Using Eq.~\eref{eqn12}, the  vectors, \textbf{u} and \textbf{v} can be defined as
\begin{eqnarray}\label{eqn16}
   \textbf{u} &\equiv \{\sqrt{P(\eta_1)}\Delta_{Q}(B_1|\eta_1), \sqrt{P(\eta_2)}\Delta_{Q}(B_1|\eta_2), ...\} \nonumber \\
   \textbf{v} &\equiv \{\sqrt{P(\eta_1)}\Delta_{Q}(B_2|\eta_1), \sqrt{P(\eta_2)}\Delta_{Q}(B_2|\eta_2), ...\}.
\end{eqnarray}
From Eq.~\eref{eqn12} and by making comparison with Eq.~\eref{eqn16}, we have, $\Delta^{2}_{min}B_{1} \geq ||\textbf{u}||_{2}^{2}$ and $\Delta^{2}_{min}B_{2} \geq ||\textbf{v}||_{2}^{2}$. Hence, we employ Eq.~\eref{eqn15} for the two outcomes $B_{1}$ and $B_{2}$ to obtain a lower bound for the product of variances $\Delta^{2}_{min}B_{1}$ and $\Delta^{2}_{min}B_{2}$ which is given by
\begin{eqnarray}\label{eqn17}
   \Delta^{2}_{min}B_{1}\Delta^{2}_{min}B_{2} &\geq ||\textbf{u}||^{2}_{2}||\textbf{v}||^{2}_{2} \geq |\langle\textbf{u},\textbf{v}\rangle|^{2} \nonumber
   \\ &\geq \left|\sum_{\eta}P(\eta)\Delta_{Q}(B_{1}|\eta)\Delta_{Q}(B_{2}|\eta)\right|^{2}.
\end{eqnarray}
We define the vector $\textbf{w}$ using the inequality Eq.~\eref{eqn17} as follows,
\begin{eqnarray}\label{eqn18}
\textbf{w} \equiv \{P(\eta_{1})\Delta_{Q}(B_{1}|\eta_{1})\Delta_{Q}(B_{2}|\eta_{1}), P(\eta_{2})\Delta_{Q}(B_{1}|\eta_{2})\Delta_{Q}(B_{2}|\eta_{2}),\nonumber\\
P(\eta_{3})\Delta_{Q}(B_{1}|\eta_{3})\Delta_{Q}(B_{2}|\eta_{3}),.......\}
\end{eqnarray}
By  utilizing Eq.~\eref{eqn18} the product of the variances in Eq.~\eref{eqn17} can be written in the following form as
\begin{eqnarray}\label{eqn19}
\Delta^{2}_{min}B_{1}\Delta^{2}_{min}B_{2}  \geq \sum_{\eta}P(\eta)^{2}\Delta_{Q}^{2}(B_{1}|\eta)\Delta_{Q}^{2}(B_{2}|\eta)
\end{eqnarray}
Using Eq.~\eref{eqn14} the above inequality can be written as,
\begin{eqnarray}\label{eqn20}
    \Delta^{2}_{min}B_{1}\Delta^{2}_{min}B_{2} \geq
    \frac{1}{4}\sum_{\eta}P(\eta)^{2}[|\langle [B_{1},B_{2}] \rangle_{\eta}|^{2} + \\ \hspace{4.5cm} \left(
\langle\{B_{1},B_{2}\}\rangle - 2\langle B_{1}\rangle \langle B_{2}\rangle \right)^{2}_{\eta}] \nonumber
\end{eqnarray}
where, $[\hat{B}_{1},\hat{B}_{2}]$ is the commutator and $\{\hat{B}_{1},\hat{B}_{2}\}$ is the anticommutator of the operators $\hat{B}_{1}$ and $\hat{B}_{2}$ respectively, $\langle B_{i} \rangle_{\eta}$ is the mean w.r.t the probability distribution g $P_{Q}(B_{i}|\eta)$. Using the properties of convex functions and Jensen's inequality, we have $\sum_{\alpha}P(\alpha)|\alpha|  \geq  |\sum_{\alpha}P(\alpha)\alpha|$, for a given probability distribution $P(\alpha)$, where $\alpha$ is a random variable. As a result, the RHS of Eq.~\eref{eqn20} in terms of inferred-variances and averages can be written as (refer to Appendix for the derivation)
\begin{eqnarray}\label{eqn21}
    \Delta^{2}_{min}B_{1}\Delta^{2}_{min}B_{2}  \geq \frac{1}{4}\left(\left|\langle [B_{1},B_{2}]\rangle\right|_{inf}\right)^{2} \\ \hspace{3.3cm} + \left(\frac{1}{2}\langle \{B_{1},B_{2}\} \rangle_{inf} - \left(\langle B_{1}\rangle \langle B_{2} \rangle\right)_{inf}\right)^{2} \nonumber 
\end{eqnarray}
From Eq.~\eref{eqn9}, using the condition $\Delta^{2}_{inf}B \geq \Delta^{2}_{min}B$ and utilizing Eq.~\eref{eqn9} in Eq.~\eref{eqn21}, we have the EPR-steering criterion based on SRUR which can be written as follows
\begin{eqnarray}\label{eqn22}
    \Delta^{2}_{inf}B_{1}\Delta^{2}_{inf}B_{2}  \geq \frac{1}{4}\left(\left|\langle [B_{1},B_{2}]\rangle\right|_{inf}\right)^{2} \\ \hspace{3.3cm} + \left(\frac{1}{2} \left\langle \{B_{1},B_{2}\} \right\rangle_{inf} - \left( \langle B_{1}\rangle \langle B_{2} \rangle \right)_{inf}\right)^{2}. \nonumber 
\end{eqnarray}
\section{Violation of our EPR-Steering criterion}
The family of isotropic states in $C_{d}\otimes C_{d}$, parameterized by $p \in \mathbb{R}$, is given by
\begin{eqnarray}\label{eqn23}
    \rho^{p}_{d} = \left(1-p\right)\frac{\mathbb{\textbf{I}}}{d^{2}} + p|\Psi_{+}\rangle\langle\Psi_{+}|
\end{eqnarray}
where, $0\leq p \leq1$, $|\Psi_{+}\rangle = \sum_{i=1}^{d} \frac{|ii\rangle}{\sqrt{d}}$ and $\mathbb{\textbf{I}}$ is the Identity operator. Here, we calculate the steerability of isotropic states for dimension $d = 2$ and $d =3$ using Eq.\eref{eqn22}.

For dimension $d=2$, two qubit isotropic state, our choice for Bob's measurement operators, $\hat{B}_{1}$ and $\hat{B}_{2}$ are spin half observables, i.e., $\hat{B}_{1} = \hat{S}_{B_x}$,$\hat{B}_{2} = \hat{S}_{B_y}$,$\hat{B}_{3} = \hat{S}_{B_z}$ with their corresponding outcomes, $S_{B_{x}}$, $S_{B_{y}}$, and $S_{B_{z}}$, respectively. We obtain a steering inequality for two-qubit isotropic state. Therefore, Eq.~\eref{eqn22} for spin observables can be written as
\begin{eqnarray}\label{eqn24}
\Delta^{2}_{inf}S_{B_x}\Delta^{2}_{inf}S_{B_z} \geq \frac{1}{4}|\langle [S_{B_{x}},S_{B_{z}}]\rangle|^{2}_{inf} +  \\ \hspace{1.9cm} \left(\frac{\langle \{S_{B_x}, S_{B_z}\}\rangle_{inf}}{2}  - (\langle S_{B_x}\rangle\langle S_{B_z}\rangle)_{inf} \right)^{2}\nonumber
\end{eqnarray}
where,  $\Delta^{2}_{inf} S_{B_i} = \langle(S_{B_i} - S_{B_i}^{est}(S_{A_i}))^{2}\rangle = \langle(S_{B_i} - g_{i}S_{A_i})^{2}\rangle, g_i = \frac{\langle S_{A_i} \otimes S_{B_i}\rangle}{\langle S_{A_i}^{2}\rangle}$, $i = x,y,z$. Calculation of inferred-variance and inferred-mean for two-qubit isotropic state gives $\Delta^{2}_{inf} S_{B_x} = \Delta^{2}_{inf} S_{B_z} = \frac{1}{4}(1 - p^{2})$, $\langle\{S_{B_x}, S_{B_z}\}\rangle = 0$, $(\langle S_{B_x}\rangle^{2})_{inf} = (\langle S_{B_y}\rangle^{2})_{inf} = (\langle S_{B_z}\rangle^{2})_{inf} = \frac{p^2}{4}$, $(\langle S_{B_x}\rangle\langle S_{B_z}\rangle)_{inf} = \frac{(1-2\sqrt{2})p^{2}-1}{16}$. Using these values in Eq.~\eref{eqn22}, we obtain the following condition that satisfies the inequality:
 \begin{eqnarray}\label{eqn25}
     p \leq 0.56.
 \end{eqnarray}
 Violation of Eq.~\eref{eqn23}, i.e. $p > 0.56$, will detect steerable two qubit isotropic states giving a better bound compared to the result of \cite{cavalcanti2009experimental}, which demonsrates steerabilty of two-qubit Werner states for $p > 0.61$, using Heisenberg uncertainty relation.  The two-qubit Werner states also show the same steerability condition \cite{wiseman2007steering}. The optimal condition for steerability of two qubit isotropic state for an infinitely large number of measurements is $p > 0.50$ \cite{wiseman2007steering}.

For dimension $d = 3$, two qutrit isotropic state, our choice of Bob's operators $\hat{B}_{1}$ and $\hat{B}_{2}$ following the commutation relation $[\hat{B}_{1},\hat{B}_{2}]=\iota\hat{B}_{3}$, given as 
 \begin{eqnarray}
 \hspace{0.5cm}B_{1} = 
     \begin{pmatrix}
         1 & 0 & 0 \\
         0 & -1 & 0 \\
         0 & 0 & 0
     \end{pmatrix}
\hspace{0.5cm}B_{2} = 
     \begin{pmatrix}
         0 & \frac{1}{\sqrt{2}} & 0 \\
         \frac{1}{\sqrt{2}} & 0 & 0 \\
         0 & 0 & 0
     \end{pmatrix}\nonumber
 \end{eqnarray}
\begin{eqnarray}
 \hspace{2.2cm}B_{3} = 
     \begin{pmatrix}
         0 & -\iota \sqrt{2} & 0 \\
         \iota \sqrt{2} & 0 & 0 \\
         0 & 0 & 0
     \end{pmatrix}
     \nonumber
\end{eqnarray}
We calculate the inferred variances and inferred means of the above operators gives $\Delta_{inf}^{2}B_{1} = \frac{2}{3}(1-p^{2}), \Delta_{inf}^{2}B_{2} = \frac{1}{3}(1-p^{2})$, $|\langle [B_{1},B_{2}] \rangle|^{2}_{inf} = \frac{p^2}{27}$, $\langle\{ B_{1},B_{2} \}\rangle_{inf} = 0$, ($\langle \hat{B}_{1} \rangle^{2})_{inf} =  \frac{2p^2}{27}$, $(\langle \hat{B}_{2} \rangle^{2})_{inf} = \frac{p^{2}}{27}$ and ($\langle B_1 \rangle \langle B_2 \rangle)_{inf} = 
-\frac{p^2}{36}$ and obtain the following steering condition by using Eq.~\eref{eqn20}
 \begin{eqnarray}
     p \leq 0.900
 \end{eqnarray}
Violation of  the above condition i.e. $p > 0.900$, will detect steerable two qutrit isotropic states. giving a better bound compared to the result of \cite{cavalcanti2009experimental}, which demonsrates steerabilty of two-qubit Werner states for $p > 0.903$, using Heisenberg uncertainty relation \cite{wiseman2007steering}. The optimal condition for steerability of two qutrit isotropic state is $p > 0.416$ for infinitely large number of measurements.

An important point to note is that our the steering bound for qutrit systems using our criterion is not a very strict bound since it involves few measurements. The experimental detection of steerable states depends crucially on the choice of measurements. Hence, the quest for an optimal choice of measurements to detect steerability of a state is a huge challenge. With an optimal choice, better bounds can be achieved using our criterion. It is also required to do larger number of measurements for fine graining of steering bounds. Recently, a covariance matrix method for arbitrary $N$ number of observables is developed. It involves covariances of observables and utilized for stronger detection of entanglement in bipartite high dimensional states  \cite{Tripathi_2020}. This method is a generalization of SRUR which involves only two observables. One of the major advantages of the covariance matrix criterion is that there is no need of knowledge about the type of states or the correlations in order to choose the optimal set operators for the detection of entanglement \cite{Tripathi_2020}. Since SRUR is a special case of the covariance matrix criterion, the LHS model and Reid's formalism can be incorporated to obtain inferred variances and expectation values of operators to formulate a complete inferred covariance matrix criterion for the detection of steering for arbitrary set of observables which could be used to capture stronger and optimal steering bounds of quantum states.

\section{Conclusion}
We utilized the Local Hidden State (LHS) model and Reid's criterion in Schr\"odinger-Robertson Uncertainty Relation (SRUR) and obtained the EPR-steering criterion for bipartite systems, which can be experimentally verified. We then used our EPR-Steering criterion to detect steerability for dimensions, $d=2$ and $d=3$ systems. We observed that  with our choice of observables, the steerability of the states could be detected, however the bounds can be further improved by selecting optimal measurement operators. 
Hence, the construction of operators which would detect steerable states optimally is a major challenge. Adopting an LHS model and Reid criterion one could obtain a completely inferred variance based steering criterion using variance based uncertainty relations, sum based uncertainty relations generalized for arbitrary number of operators involving higher order moments of the operators. One of the future aspects of our criterion is to look for steering bounds in continuous variable systems. Furthermore, this criterion can be implemented using measurements corresponding to positive operator-valued measures (POVM). Many works show the correspondence between joint measurability and steering. Therefore, it would be interesting to look for the correspondence between steering and uncertainty relations involving incompatible POVMs.

\section{Data availability}
All data generated or analysed during this study are included in this published article

\section{Acknowledgments}
We would like to thank Dr. Debashis Saha and Mr. Sumit Mukherjee for numerous enlightening discussions. PKP acknowledges the support from DST, India, through Grant No. DST/ICPS/QuST/Theme-1/2019/2020-21/01

\appendix
\section*{Appendix}\label{append}
\setcounter{section}{1}
Consider the inequality developed in Eq.~\eref{eqn20}. For Bob's two outcomes $B_1$ and $B_2$, corresponding to measurement operators $\hat{B}_{1}$ and $\hat{B}_{2}$ we have the following expression
\begin{eqnarray}\label{a1}
     \Delta^{2}_{min}B_{1}\Delta^{2}_{min}B_{2} \geq 
     \frac{1}{4}\sum_{\eta}P(\eta)^{2}|\langle [B_{1},B_{2}]\rangle_{\eta}|^{2} + \\ 
     \hspace{3.2cm}\sum_{\eta}P(\eta)^{2}\left(\frac{\langle\{B_{1}, B_{2}\}\rangle}{2} - \langle B_{1}\rangle \langle B_{2}\rangle \right)^{2}_{\eta}.\nonumber
\end{eqnarray}
The terms on RHS of the Eq.~\eref{a1}, $\sum_{\eta}P(\eta)|\langle [B_{1},B_{2}]\rangle_{\eta}|$ and $\sum_{\eta}P(\eta)\left(\frac{\langle\{B_{1}, B_{2}\}\rangle}{2} - \langle B_{1}\rangle \langle B_{2}\rangle \right)_{\eta}$ can be defined as vectors with their inner products written as follows
\begin{eqnarray}\label{a2}
\geq \frac{1}{4}\sum_{\eta}P(\eta)|\langle B_3\rangle_{\eta}| \hspace{0.05cm} \textbf{.} \hspace{0.05cm} \sum_{\eta'}P(\eta')|\langle B_3\rangle_{\eta'}|\\  + \sum_{\eta}P(\eta)\left(\frac{\langle B_{4} \rangle}{2} - \langle B_{1}\rangle \langle B_{2}\rangle \right)_{\eta} \nonumber \\ \hspace{3.8cm} \textbf{.} \hspace{0.05cm} \sum_{\eta'}P(\eta')\left(\frac{\langle B_{4}\rangle}{2} - \langle B_{1}\rangle \langle B_{2}\rangle \right)_{\eta'} \nonumber
\end{eqnarray}
where, $[\hat{B}_{1},\hat{B}_{2}] = \iota\hat{B}_{3}$ and $\{\hat{B}_{1},\hat{B}_{2}\} = \hat{B}_{4}$. Here, we witness one of the important applications of LHV, i.e., the sufficient condition of reality. It is very important to note that in the LHS model, the choice of the operators $\{\hat{A_1},\hat{A_2},\hat{A_3}\}$ for Alice is completely arbitrary. In this model, the operators are allowed to  depend arbitrarily on the hidden variable $\eta$ and hence play no role in the LHS model of Bob. For all the measurement outcomes $A_{i}$, corresponding to the operators $\hat{A}_{i}$, that exhaust Alice's measurement set of operators, the inequality \eref{a2} can be written in the following form
\begin{eqnarray}\label{a3}
\geq \frac{1}{4}\sum_{\eta,A_3}P(\eta,A_3)|\langle B_3\rangle_{\eta}|\hspace{0.05cm} \textbf{.} \hspace{0.05cm} \sum_{\eta',A_3}P(\eta',A_3)|\langle B_3\rangle_{\eta'}| \\ + \left(\frac{1}{2}\sum_{\eta,A_4}P(\eta,A_4)\langle B_{4} \rangle_{\eta} - \sum_{\eta} P(\eta) (\langle B_{1} \rangle \langle B_{2} \rangle)_{\eta} \right)\hspace{0.05cm} \nonumber \\ \hspace{1.7cm} \textbf{.} \left(\frac{1}{2}\sum_{\eta,A_4}P(\eta,A_4)\langle B_{4} \rangle_{\eta'} - \sum_{\eta'} P(\eta') (\langle B_{1} \rangle \langle B_{2} \rangle)_{\eta'} \right) \nonumber
\end{eqnarray}
where $[\hat{A}_1,\hat{B}_2] = \iota \hat{A}_3$, $\{\hat{A}_1,\hat{A}_2\} = \hat{A}_4$, all the $P$'s are the joint probability distribution for different outcomes $A_{i}$ and classical random variable $\eta$. $P(A_{i})$ for $i = 1,2,3,4$ corresponds to the outcomes $A_{i}$ of the measurement observables $\hat{A}_{i}$.
Here, we utilize the Bayes rule and the marginal distribution $P(A_{i}) = \sum_{\eta}P(\eta,A_{i}) = \sum_{\eta}P(\eta|A_{i})P(A_{i})$ where, $A_{i}$ and $A_{j}$ are the outcomes corresponding to Alice's different measurements $\hat{A_{i}}$ and $\hat{A_{j}}$ respectively. $\eta$ is the local hidden variable correlated to the local hidden state $\sigma_{\eta}$. For a real variable $u$, $|u|^{2}$ and $u^{2}$ are convex functions. Hence, by Jensen's inequality, $\sum_{u}P(u)|u| \geq |\sum_{u}P(u)u|$. Therefore, expression \eref{a3} is lower bounded as follows
\begin{eqnarray}\label{a4}
\geq \frac{1}{4}\sum_{A_3}P(A_3)\left|\sum_{\eta}P(\eta|A_3)\langle B_3\rangle_{\eta}\right| \\ \hspace{3.9cm} \hspace{0.05cm} \textbf{.} \hspace{0.05cm} \sum_{A_3}P(A_3)\left|\sum_{\eta'}P(\eta'|A_3)\langle B_3\rangle_{\eta'}\right| \nonumber\\ + \left(\frac{1}{2}\sum_{A_4}P(A_4)\sum_{\eta}P(\eta|A_4)\langle B_{4} \rangle_{\eta} - \sum_{\eta} P(\eta) (\langle B_{1} \rangle \langle B_{2} \rangle)_{\eta} \right)\hspace{0.05cm} \nonumber \\ \hspace{0.3cm} \textbf{.} \left(\frac{1}{2}\sum_{A_4}P(A_4)\sum_{\eta'}P(\eta'|A_4)\langle B_{4} \rangle_{\eta'} - \sum_{\eta'} P(\eta') (\langle B_{1} \rangle \langle B_{2} \rangle)_{\eta'} \right) \nonumber
\end{eqnarray}
We consider the last term of the expression \eref{a4}, derive and explain how product of the expectation values of the outcomes $B_{1}$ and $B_{2}$ of Bob corresponding to his measurement operators $\hat{B}_{1}$ and $\hat{B}_{2}$ can be jointly inferred from Alice's three measurements as follows
\begin{eqnarray}\label{a5}
    \sum_{\lambda}P(\lambda)(\langle B_{1}\rangle_ \langle B_{2}\rangle)_{\lambda} = \\ \hspace{2.7cm} \frac{1}{2}\sum_{\lambda}P(\lambda)[\langle B_{1}\rangle_{\lambda}^{2} + \langle B_{2} \rangle_{\lambda}^{2} - \langle B_{1} - B_{2}\rangle^{2}_{\lambda}]. \nonumber
\end{eqnarray}
The inferred value of the product of the expectation values of Bob's outcomes $B_1$ and $B_2$ corresponding to his measurements $\hat{B}_{1}$ and $\hat{B}_{2}$ i.e., $\langle B_{1} \rangle \langle B_{2} \rangle$ cannot be obtained from a local hidden state for Bob from the outcomes $A_1$ and $A_2$ of Alice's two measurements $\hat{A}_{1}$ and $\hat{A}_{2}$. In the operational definition of steering Alice sends a local hidden state $\sigma^{A}_{a}$ to Bob, by picking from an ensemble $\textnormal{R} = \sum_{\eta}p_{\eta}\hat{\sigma}_{\eta}$ and doing a stochastic map from the LHV $\eta$ to the result $a$ of the measurement $\hat{A}$ in each run of the experiment. Alice randomly chooses any one of the results of the measurement $A$ corresponding to which, she wants Bob's state to collapse into. For the inferred value of the product of the expectation value of two operators Alice has to encode two outcomes corresponding to her choice of two incompatible measurements and since LHS is a quantum state Alice cannot perform this due to the incompatibility of the measurement operators. Hence, Alice cannot obtain the inferred product from the two measurements. However, Alice can infer the product of the expectation values by performing three different incompatible measurements $\hat{A}_{1}$ and $\hat{A}_{2}$ and $\hat{A}_{1} - \hat{A}_{2}$ respectively. Therefore the inferred value of the product of the expectation value is given as
\begin{eqnarray}\label{a6}
    2(\langle B_{1} \rangle \langle B_{2} \rangle)_{inf} = \sum_{A_1,\lambda} P(A_1,\lambda) \langle B_{1} \rangle^{2}_{\lambda} \\ \hspace{2.7cm}+ \sum_{A_2,\lambda} P(A_2,\lambda) \langle B_{2} \rangle^{2}_{\lambda} - \sum_{A_0,\lambda} P(A_0,\lambda) \langle B_{0} \rangle^{2}_{\lambda} \nonumber
\end{eqnarray}
where we denote $B_0$ as outcome corresponding to Bob's measurement $\hat{B}_{0} = \hat{B}_{1} - \hat{B}_{2}$ and and $A_0$ corresponding to Alice's measurement $\hat{A}_{0} = \hat{A}_{1} - \hat{A}_{2}$ respectively. The Eq.\eref{a6} is given as follows
\begin{eqnarray}\label{a7}
    2(\langle B_{1} \rangle \langle B_{2} \rangle)_{inf} = \sum_{A_1} P(A_1) \sum_{\lambda} P(\lambda|A_1) \langle B_{1} \rangle^{2}_{\lambda} +\\ \sum_{A_2} P(A_2)\sum_{\lambda}P(\lambda|A_2) \langle B_{2} \rangle^{2}_{\lambda} - \sum_{\lambda} P(A_0) \sum_{\lambda}P(\lambda|A_0) \langle B_{0} \rangle^{2}_{\lambda}.\nonumber
\end{eqnarray}
In the above Eq.\eref{a7}, the expectation values of Bob's measurement outcomes conditioned on Alice's measurement outcomes is written as
\begin{eqnarray}\label{a8}
    2(\langle B_{1} \rangle \langle B_{2} \rangle)_{inf} = \sum_{A_1} P(A_1) \langle B_{1} \rangle^{2}_{A_1} + \sum_{A_2} P(A_2) \langle B_{2} \rangle^{2}_{A_2} \\ \hspace{5cm}- \sum_{A_0} P(A_0) \langle B_{0} \rangle^{2}_{A_0} \nonumber \\
    \hspace{2cm} = \left(\langle B_{1} \rangle^{2}\right)_{inf} + \left(\langle B_{2} \rangle^{2}\right)_{inf} - \left(\langle B_{0} \rangle^{2}\right)_{inf} \nonumber \\
    \hspace{2cm} = \left(\langle B_{1} \rangle^{2}\right)_{inf} + \left(\langle B_{2} \rangle^{2}\right)_{inf} - \left(\langle B_{1}  - B_{2} \rangle^{2}\right)_{inf}. \nonumber
\end{eqnarray}
We use the derivation of Eq.\eref{a7} in Eq.\eref{a4} and obtain a complete inferred expression on the RHS of the Eq.\eref{a1}
\begin{eqnarray}\label{a9}
\geq \frac{1}{4}\sum_{A_3}P(A_3)|\langle B_3\rangle_{A_3} \textbf{.} |  \sum_{A_3}P(A_3)|\langle B_3\rangle_{A_3}|  \\+ \left(\frac{1}{2}\sum_{A_4}P(A_4)\langle B_{4} \rangle_{A_4} - \left(\langle B_{1}\rangle \langle B_{2} \rangle\right)_{inf}\right) \nonumber \\ \hspace{3.4cm}\textbf{.} \left(\sum_{A_4}P(A_4) \langle B_4\rangle_{A_4} - \left(\langle B_{1}\rangle \langle B_{2} \rangle\right)_{inf} \right) \nonumber
\end{eqnarray}
The above expression can be further simplified into the following form
\begin{eqnarray}\label{a10}
    \geq \frac{1}{4}|\langle B_3\rangle|_{inf}|\textbf{.}
    \langle B_3\rangle|_{inf} + \left(\frac{1}{2}\langle B_{4} \rangle_{inf} - \left(\langle B_{1}\rangle \langle B_{2} \rangle\right)_{inf}\right) \\ \hspace{4.2cm}\textbf{.} \left( \frac{1}{2}\langle B_4\rangle_{inf} - \left(\langle B_{1}\rangle \langle B_{2} \rangle\right)_{inf} \right) \nonumber
\end{eqnarray}
\begin{eqnarray}
     \geq \frac{1}{4}\left(\left|\langle B_3\rangle\right|_{inf}\right)^{2} + \left(\frac{1}{2}\langle B_{4} \rangle_{inf} - \left(\langle B_{1}\rangle \langle B_{2} \rangle\right)_{inf}\right)^{2}\nonumber
\end{eqnarray}
We obtain the complete expression involving the variances of the operators $\hat{B}_1$ and $\hat{B}_2$ which is given as
\begin{eqnarray}\label{a11}
     \Delta^{2}_{min}B_{1}\Delta^{2}_{min}B_{2} \geq \frac{1}{4}\left(\left|\langle B_3\rangle\right|_{inf}\right)^{2} + \left(\frac{1}{2}\langle B_{4} \rangle_{inf} - \left( \langle B_{1}\rangle \langle B_{2} \rangle \right)_{inf}\right)^{2} \nonumber \\
\end{eqnarray}
\section*{References}
\bibliography{main}
\end{document}